# New Aspects of the Temperature – Magnetic Field Phase Diagram of $CeB_6$

R. G. Goodrich[1], David P. Young[1], Donavan Hall[2], Z. Fisk[2], N. Harrison[3], J. Betts[3], Albert Migliori[3], F. M. Woodward[4] and J. W. Lynn[4]

[1]Department of Physics and Astronomy, Louisiana State University, Baton Rouge, LA 70803
[2]National High Magnetic Field Laboratory, Florida State University, Tallahassee, FL 32306
[3]National High Magnetic Field Laboratory, Los Alamos National Laboratory, Los Alamos, NM 87545
[4]NIST Center for Neutron Research, National Institute of Standards and Technology, Gaithersburg, MD 20899-8562

**Abstract**

We have measured the magnetic field dependence of the paramagnetic to the field-induced high temperature antiferroquadrupolar magnetically ordered phase transition in $CeB_6$ from 0 to 60 T using a variety of techniques. It is found that the field-dependent phase separation line becomes re-entrant above 35 T and below 10 K. Measurements of resonant ultra-sound, specific heat and neutron diffraction show conclusively that the zero-field temperature-dependent phase transition is to a state with no ordered dipole moments, but with second order transition signatures in both the sound attenuation and specific heat.

The nature of cooperative ordering in highly correlated electron systems continues to be a central topic of fundamental interest. In the past decade cerium hexaboride ($CeB_6$) and related materials have been the focus of many studies of their electronic, thermal, and magnetic properties to investigate the delicate balance between the Kondo-lattice ground state of these highly correlated electron systems, and various states with long range magnetic order.[1] $CeB_6$ is a prototype system because all the interesting properties arise from a single *4f* electron on the Ce ion that hybridizes with the conduction electrons, giving rise to heavy fermion (HF) behavior. Three different phases so far have been identified in this material. Phase I occurs at high temperatures (≥ 10K), where $CeB_6$ is paramagnetic and exhibits the Kondo effect (electrical resistivity increasing logarithmically with decreasing temperature)[2]. In zero applied field an ordered state identified as antiferroquadrupolar (AFQ) order develops below $T_q$=3.3 K (Phase II), while conventional dipolar antiferromagnetic order develops below $T_N$=2.3 K (Phase III).[3] As a function of applied magnetic field, Phase III has been reported to exhibit three different orderings of the dipole moments, with phase boundaries that come together at 2.3 K in zero field, while the phase boundary for the AFQ order has recently been found to increase with increasing field with no indication of re-entrant behavior up to 30 T (see Fig. 1).[4] In this paper we present several types of measurements on high-quality single crystals of $CeB_6$. First, we have extended the phase boundary between Phase I and II to 60 T, and observe that above 35 T Phase I becomes re-entrant, in agreement with expectations of theory.[5] Second, we show that a weak second order phase transition does occur at zero field between Phases I and II as a function of temperature. Third, we show from neutron diffraction that in zero applied magnetic field no magnetic dipole ordering exists in Phase II. Fourth, we identify a new zero-field second order phase transition at 1.6 K, inside the accepted Phase III regime.

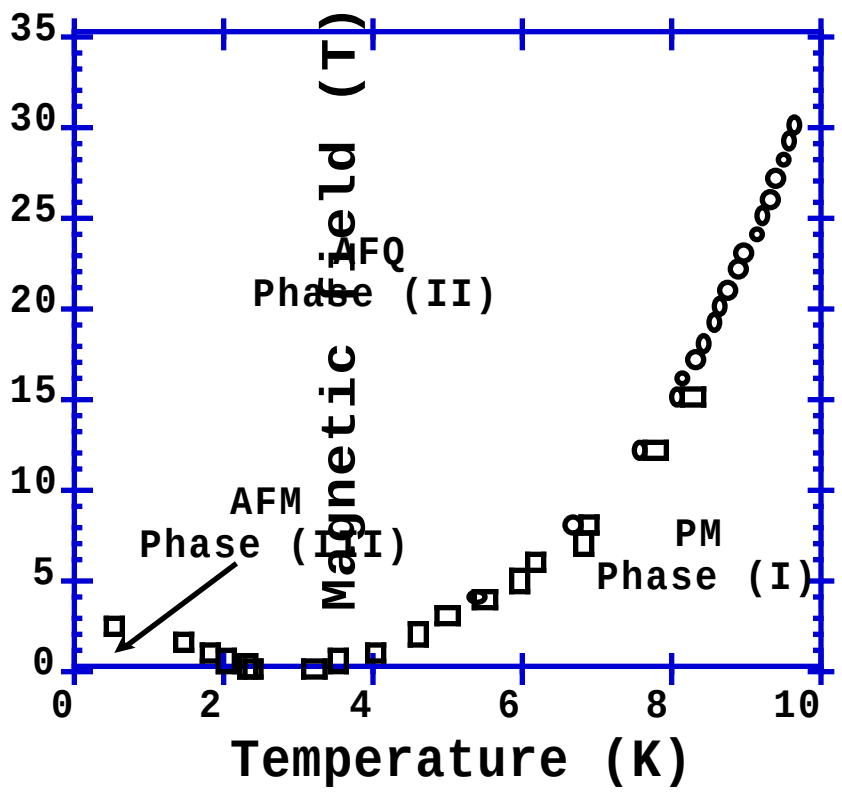

Figure 1. Phase diagram for $CeB_6$ determined previously, showing three main phases. At zero field there are two magnetic ordering temperatures; the quadrupolar ordering at $T_q$=3.3 K, and the Néel temperature $T_N$=2.3 K.

Cerium hexaboride is one of several rare earth hexaborides that crystallize in the primitive cubic structure with the rare earth ions at the cube center and boron octahedra at the cube corners. The cubic crystal field due to the six boron atoms in $CeB_6$ splits the single electron *4f* 6-fold degenerate $^2F_{5/2}$ level into a 2-fold degenerate $\Gamma_7$ and a 4-fold degenerate $\Gamma_8$ level[6].

It has been shown that in $CeB_6$ the $\Gamma_8$ is the lowest energy state, and the splitting between the $\Gamma_7$ and the $\Gamma_8$ levels is on the order of 530 K[7]. The $\Gamma_8$ symmetry of the f electron on Ce allows not only a magnetic dipole moment, but higher order moments, including orbital electric and magnetic quadrupole moments, as well as a magnetic octapole moment.

For the high-field phase boundary determination we have made measurements using two different techniques. First, temperature dependent cantilever magnetometer measurements of the sample magnetization between 25 T and 45 T were made in steady fields using the hybrid superconducting-plus-resistive magnet at the National High Magnetic Field Laboratory (NHMFL) in Tallahassee, FL. This method of measurement previously has been described in Ref. 4. Second, we made constant temperature susceptibility measurements in pulsed fields to 60 T at the NHMFL, Los Alamos. For these latter measurements we placed the sample in a balanced pickup coil to measure the change in susceptibility of the sample as a function of field. The results of these measurements, along with our previous data, are shown in Fig 2. In the inset we show a quadratic fit to the data showing that, if the fit were to persist, the zero temperature transition would occur at 80 T.

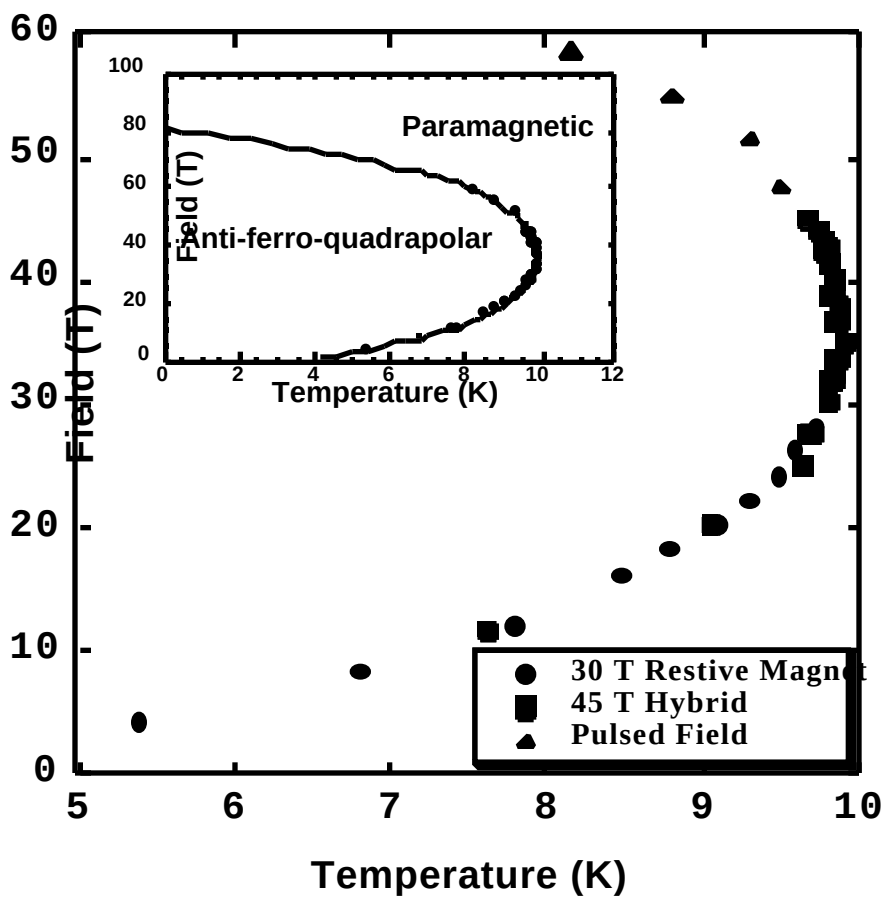

Figure 2. Quadrupolar ordering temperature ($T_q$) vs. field from previously published results[1] plotted with the current data that now extends to 60 T. Inset: Quadratic fit to the data suggesting the zero-temperature transition would occur near 80 T.

Uimin[8] has described the shape of $T_q$ vs. H as arising from competing AFQ patterns near the ordering temperature. These fluctuations are suppressed by an applied magnetic field. Uimin's model predicts three important characteristics of the AFQ-PM phase diagram: (1) that $T_q$ vs. H increases linearly at low applied fields, (2) that the AFQ-PM phase line is anisotropic in the T-H plane, and (3) that $T_q$ vs. H decreases and goes to zero at sufficiently high fields. Based on previously published data Uimin estimated the lower limit field for the re-entrance of $T_q$ vs. H as approximately 25 - 30 T, yielding an $H(T_q = 0)$ approaching 80 T. The measurements reported here do show re-entrance above 30 T and project to an $H(T_q = 0)$ at 80 T. Uimin points out that his estimate of $H(T_q = 0)$ does not take into account the Kondo effect, but ignoring the Kondo effect is a valid approximation at fields H >> 2 T where the *f*-electrons are localized and the Kondo interaction is therefore weak.

The ordering in Phase II previously has been studied in the presence of applied magnetic fields by neutron diffraction. Quadrupolar order is not observed directly with neutrons, but a magnetic field induces magnetic dipole moments on the periodic structure of ordered electric quadrupole moments[9]. The corresponding wave vector $k_0$ = [1/2, 1/2, 1/2] was observed in neutron diffraction[9], and the ordering in Phase II was proposed to be that of electric quadrupole moments, requiring a splitting of the four-fold degenerate $\Gamma_8$ ground state into two doublets. Several models have been given for this splitting, including a dynamic Jahn-Teller effect involving acoustic phonons or a hybridization-mediated anisotropic coupling of the 4f wave functions to the p-like boron or 5d-type cerium wave functions[7]. In an early paper Ohkawa[10] proposed that indirect exchange interactions between pairs of Ce atoms would produce a splitting of the four-fold degenerate level into (4 x 4) sixteen levels, split into a group of two triplets and a group consisting of a singlet plus a nine-fold degenerate level. More recently, an alternate interpretation of the neutron scattering results was given by Uimin[10] in which the low temperature frequency shift of the $\Gamma_7$ - $\Gamma_8$ splitting was interpreted as arising from collective modes caused by the orbital degrees of freedom. It should be noted that muon spin rotation measurements in zero applied magnetic field yield a different magnetic structure for $CeB_6$ for both Phase II and Phase III[11], but it does show that the exchange coupling between Ce atoms must be antiferromagnetic.

To investigate the nature of the zero-field ordering, we performed neutron diffraction measurements between 1.4 K and 4.5 K on 99% $^{11}$B-enriched $CeB_6$ at the NIST Center for Neutron Research. The sample was mounted in the [*hhl*] scattering plane with no applied field (<$10^{-4}$ T). A pyrolytic graphite monochromator and filter were employed at neutron wavelengths of 2.359 Å for BT-2 and 2.461 Å on BT-7, with relaxed angular collimations to optimize the observed intensities. Magnetic dipole moments (M1) and electric quadrupole moments (E2) have the same parity symmetry, but have opposite time reversal symmetry (M1 – odd and E2 – even). Thus the application of an external magnetic field is required to break the time reversal symmetry of M1 and allow the magnet dipole moments to co-exist with the electric quadrupole moments. Fig. 3 shows a comparison of the antiferromagnetic peak observed at the (1/4,1/4,1/4) position at 1.6 K. The asymmetry of the scattering originates from the mosaic of the crystal. The ordered antiferromagnetic moment we observe is <m>=0.26(4) $\mu_B$, in good agreement with the literature. Also shown in the figure is the scattering at the AFQ position, (1/2,1/2,1/2). No scattering was detectable at 1.6, 2.8 K, or 3.3 K, which places an upper limit of 0.03$\mu_B$, for any induced dipole moment associated with the AFQ order. Note in particular that the data taken at 2.8 K is well within the Phase II region, and the absence of a peak shows that there is no significant zero-field magnetic ordering due to coupling with quadrupolar moments, in contrast to the ordering that is readily observed in the Phase II regime for fields of 1 T and above[9].

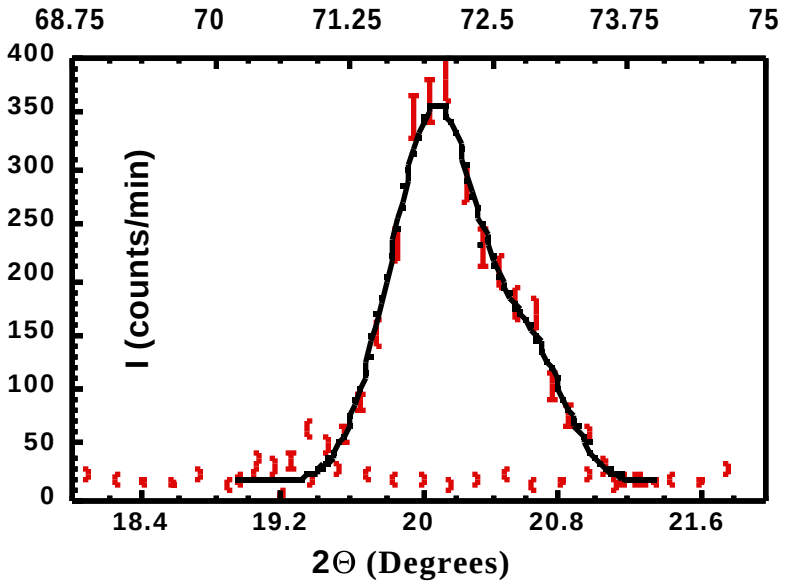

Figure 3. Direct comparison of the (1/4,1/4,1/2) antiferromagnetic peak taken at 1.6 K (upper scale) with the absence of scattering (lower scale) at the (1/2,1/2,1/2) AFQ position at 2.8 K.

In order to verify that a phase transition occurs from Phase I to Phase II, and that a further transition to Phase III exists at zero field, both resonant ultrasound (RUS) and specific heat measurements were made at the NHMFL (Los Alamos), again in fields $<10^{-4}$ T. The RUS measurements gave signals proportional to the elastic response of a single crystal of $CeB_6$ from 1.2 K to 4.2 K, with emphasis on the region near the phase transitions at 3.3 K and 2.3 K. With a resolution of 1 part in $10^6$, we observe two changes in the elastic tensor near the AFQ ($T_q$), and the AFM ($T_{AF}$) phase transitions. These changes are shown in Fig. 4a, where the transition $T_q$ at 3.3 K is observed as a change in slope in the temperature dependent data, and $T_N$ at 2.3 K shows a discontinuity with no hysteresis. These data indicate that the 3.3 K transition ($T_q$) is weakly second order, while the AFM transition ($T_N$) is a sharp second order transition. A second zero-field transition, $T_2$, is observed in Phase III near 1.6 K. In addition to the RUS results, temperature dependent specific heat results at zero field are shown in Fig. 4b. Again, the $T_q$ transition is seen to be weakly second order, and $T_N$ is sharp. To verify the thermodynamic order of $T_q$ and $T_N$, specific heat measurements were made using a pulse-relaxation technique for both increasing and decreasing temperature. The results shown in Fig. 4b clearly demonstrate the nature of the transitions with no hysteresis in either, indicating that the $T_N$ transition is also second order. Further measurements in applied fields up to 15 T showed that both transitions followed the published phase diagram.

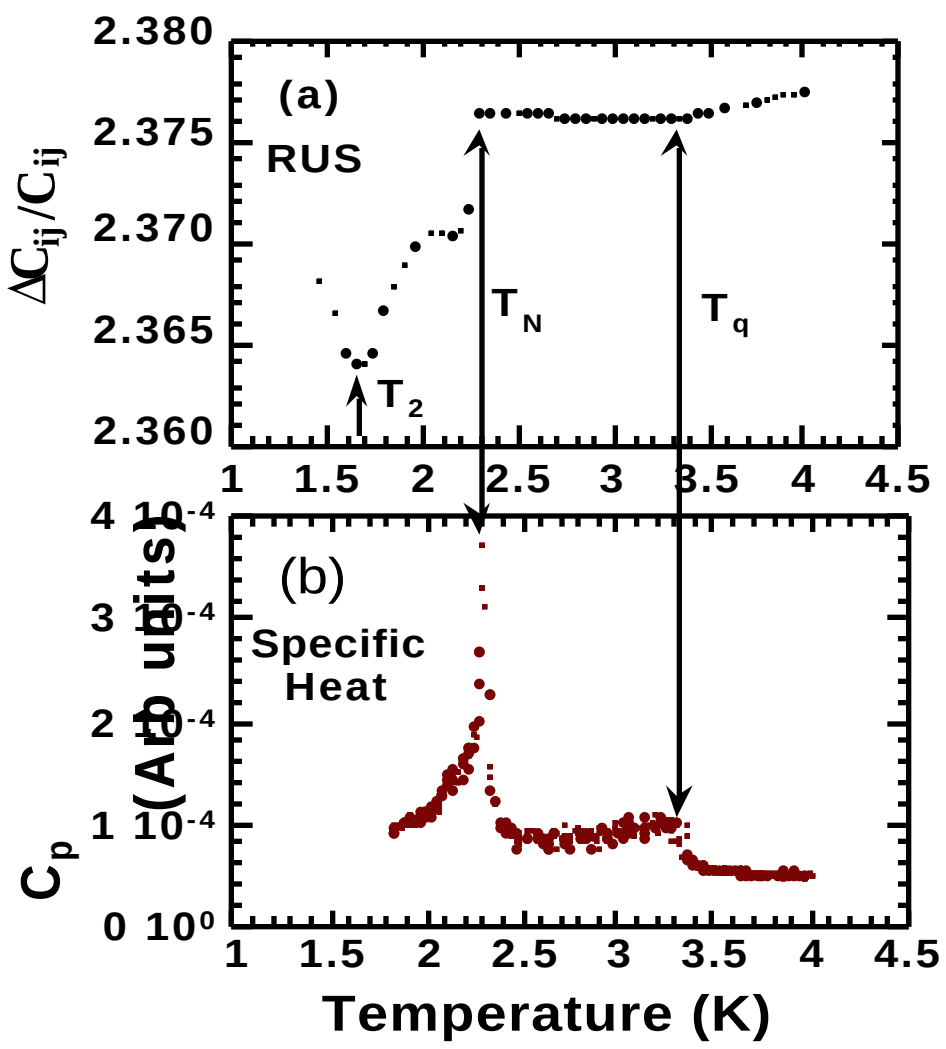

Figure 4 (a.) Relative change in the elastic constants of a single crystal of cubic $CeB_6$ as measured by RUS as a function of temperature. (b) Temperature dependence of the heat capacity of $CeB_6$. The Phase I – Phase II transition (3.3 K), although weak, occurs at zero field, and the Phase II – Phase III transition (2.3 K) is extremely sharp and second order.

For the first time the $T_q$ phase diagram of $CeB_6$ is seen to reverse direction in temperature with increasing field, and then the paramagnet phase becomes re-entrant as a function of field and temperature. All of the present data, taken together, show that Phase II in $CeB_6$ exists at zero field, but does not become magnetically ordered until the application of an external magnetic field breaks the time reversal symmetry between the dipole and quadrupole moments. We have observed that this magnetically ordered field-induced phase can be destroyed by fields only exceeding 35 T. Based on these results, any theory that predicts the destruction of Phase II below 30 T does not include either all of the effects, or it includes incorrect mechanisms. However, two of the theories presented to date, both of which are predicated on indirect exchange, predict destruction of the AFQ phase at fields > 30 T, and cannot be ruled out. Finally, when the RKKY interaction between the 4f electron spins overcomes the thermal effects, internal magnetic fields are produced that break time-reversal symmetry and allow spontaneous ordering of the magnetic moments in Phase III.

A portion of this work was performed at the National High Magnetic Field Laboratory, which is supported by NSF Cooperative Agreement No. DMR-9527035 and by the State of Florida. Two of the authors, RGG and DPY, wish to thank A. R. P. Rau and Dana Browne for several helpful conversations.